\documentclass[epj
]{svpreprint} 
\usepackage{graphicx}
\begin{document}
%
\title{Derivation of a Fundamental Diagram for Urban Traffic Flow}
\author{Dirk Helbing
}                     
%
%
\institute{ETH Zurich, UNO D11, Universit\"atstr. 41, 8092 Zurich, Switzerland}
\date{Received: date / Revised version: date}
%
\abstract{Despite the importance of urban traffic flows, there are only a few theoretical approaches to determine fundamental relationships between macroscopic traffic variables such as the traffic density, the utilization, the average velocity, and the travel time. In the past, empirical measurements have primarily been described by fit curves. Here, we derive expected fundamental relationships from a model of traffic flows at intersections, which suggest that the recently measured fundamental diagrams for urban flows can be systematically understood. In particular, this allows one to derive the average travel time and the average vehicle speed as a function of the utilization and/or the average number of delayed vehicles.
\PACS{
      {89.40.Bb}{Land transportation} \and
      {47.10.ab}{Conservation laws and constitutive relations} \and
      {51.10.+y}{Kinetic and transport theory of gases} 
     } 
} 
\maketitle
\section{Introduction}

The study of urban traffic flows has a long history (see Ref. \cite{Gazis} for an overview). 
For more than a decade now, physicists have contributed various interesting models, ranging from cellular automata \cite{Esser,Nagel1,Nagel} to fluid-dynamic approaches \cite{Hilliges,NHM}. Complementary, one should mention, for example, Refs. \cite{Cremer,Daganzo} as representatives of publications by traffic engineers, and also continuous microscopic flow models used in commercial software tools such as VISSIM.
\par
One research area in traffic physics is the transition from free to congested traffic in urban road networks 
\cite{NagaCity,Chowdhury}, which started off with the paper by Biham, Middleton and Levine \cite{BML}. Interestingly enough, the spreading of congestion seems to share some features with cascading failures \cite{Zheng2007}.
\par 
In the following, we will focus on the study of fundamental relationships between flow, utilization, and density on the one hand and the average velocity or travel time on the other hand.
Such relationships were already studied in the 60ies 
\cite{Irwin,Smock,Mosher,BPR,Soltmann,Davidson,Smeed,Over,Thomson,Wardrop,Zahavi}, but this work primarily took a phenomenological approach. Moreover, the recorded data in the regime of congested road networks scattered enormously. Thus, it was hard to fit a curve. Consequently, many different relationships were proposed. The probably most wide-spread formula
is the one published by the Bureau of Public Roads \cite{BPR}. Accordingly, the travel time ${\cal T}_i$ of an urban road section $i$ with capacity $C_i$ would follow the capacity constraint function 
\begin{equation}
 {\cal T}_i = {\cal T}_i^0 \left[ 1 + \alpha_i \left(\frac{A_i}{C_i}\right)^{\beta_i} \right] \, .
 \label{capres}
\end{equation}
(see Fig. \ref{fIG4}). Here, $A_i$ is the arrival flow in that road section and ${\cal T}_i^0$ the travel time for light traffic conditions, while $\alpha_i \approx 0.5$ and $\beta_i \approx 4$ are fit parameters. It is obvious that this formula does not diverge when the capacity $C_i$ is reached, as one may expect. This is just one of the many theoretical inconsistencies of the proposed phenomenological formulas. It is no wonder that the subject of fundamental diagrams for urban traffic has recently been taken up again \cite{Akcelik,Lum,Zhang,Tu,Daganzo2007,Geroliminis,Simdata,DagGer}. 
\par
The first measurement of an urban fundamental diagram was presented by Godfrey \cite{Godfrey}.
Ten years later, even Robert Herman and nobel prize laureate Ilya Prigogine addressed traffic flow in cities \cite{twofluid,twofluid1}. They tried to derive fundamental relationships via a statistical physics approach, which however still contained phenomenological elements. Their ``two-fluid approach'' considered moving and standing traffic and the mutual interdependencies between them. 
\par
Recently, the issue of urban gridlock was reanimated by Carlos Daganzo \cite{Daganzo2007}.
Together with Geroliminis, he presented convincing evidence for a fundamental diagram of urban traffic flow \cite{Geroliminis,Simdata}. Before, fundamental diagrams were mainly used for the study of freeway systems \cite{Review}. There, they were very useful to understand capacity effects \cite{Gazis}, different traffic states \cite{EPJBphase}, and the traffic dynamics, in particularly the propagation of shock fronts \cite{Whitham}. It is, therefore, not surprising that people are eager to find fundamental relationships for urban traffic as well. 
\par
The most recent progress is an approximation of the urban fundamental diagram by Daganzo and Geroliminis, which is based on cutting away parts of the flow-density plane \cite{DagGer}. It appears possible to construct a relationship with kinematic wave theory \cite{Eichler}. In contrast to the density-based approach by Daganzo {\it et al.}, we will pursue an alternative, utilization-based approach. This is common in queueing theory \cite{Little} and transportation planning, where formulas such as the capacity constraint function (\ref{capres}) are used.
\par
After discussing elementary relationships for cyclically signalized intersections of urban road networks in Sec. \ref{cyclic}, we will start in Sec. \ref{under} with the discussion of {\it undersaturated} traffic conditions and derive a relationship for 
the average travel time as a function of the utilization or the number of delayed vehicles. Furthermore, we will determine a formula for the average speed. In Sec. \ref{congest}, we will extend the analysis to {\it congested} road conditions, where the intersection capacity is exceeded. Afterwards, in Sec. \ref{over}, we will indicate how to deal with {\it oversaturated} networks, where the link capacity is insufficient to take up all vehicles that would like to enter a road section. Finally, Sec. \ref{sum} provides a summary and discussion. In particularly, we will address issues regarding the transfer of link-based fundamental diagrams to urban areas. We also try to connect the density-based fundamental diagram of Daganzo {\it et al.} with the utilization-based approach developed here.

\section{Elementary Relationships for Cyclically Operated Intersections}\label{cyclic}

Let us study a single intersection with a periodically operated traffic light. We shall have green phases $j$ of duration $\Delta T_j$, during which one or several of the traffic streams $i$ are served.
$\beta_{ij}$ shall be 1, if traffic stream $i$ is served by green phase $j$, otherwise $\beta_{ij} = 0$.
The setup time after phase $j$ shall require a time period $\tau_j$. It may be imagined to correspond to the time period of the amber light (although in practice, this has to be corrected for reaction times and intersection clearing times). The sum of setup times will be called the ``lost service time''
\begin{equation}
T_{\rm los} = \sum_j \tau_j \, ,
\end{equation}
while the sum of all green time periods and setup times will be called the ``cycle time''
\begin{equation}
T_{\rm cyc} = \sum_j (\Delta T_j + \tau_j) = T_{\rm los} + \sum_j \Delta T_j \, .
\label{ater}
\end{equation}
The green times are also sometimes expressed as fractions $f_j\ge 0$ of the cycle time, i.e.
\begin{equation}
\Delta T_j = f_j T_{\rm cyc} 
\label{fdef}
\end{equation}
with $\sum_j f_j < 1$. After inserting this into Eq. (\ref{ater}) and rearranging terms, we get
\begin{equation}
 T_{\rm cyc}(\{f_j\}) = \frac{T_{\rm los}}{1 - \sum_j f_j} \, ,
 \label{accordi}
\end{equation}
i.e. the cycle time is proportional to the lost service time $T_{\rm los}$.
\par
\begin{figure}[htbp]
\begin{center}
\includegraphics[width=9cm]{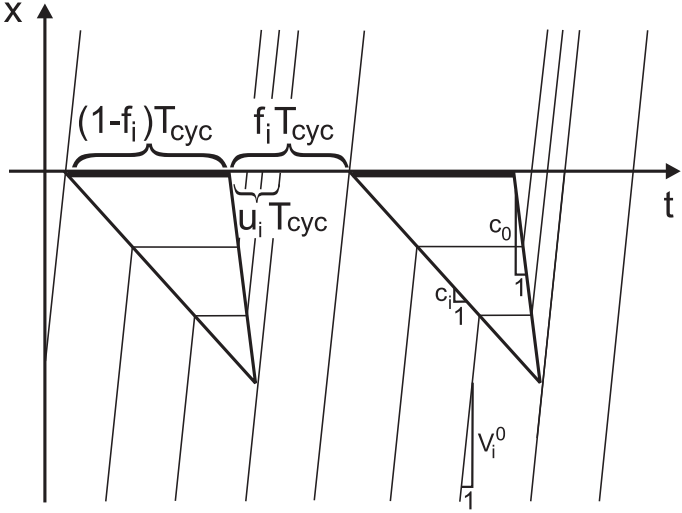}
\end{center}
\caption[]{Schematic illustration of vehicle trajectories for a traffic light, which has an amber and red time period of altogether $(1-f_i)T_{\rm cyc}$ and a green time period of $f_iT_{\rm cyc}$. Vehicles move forward at the free speed $V_i^0$ or are stopped in a vehicle queue (horizontal lines), which forms during the amber and red time period behind the traffic light (located at the $t$-axis). The speed of the upstream moving congestion front is given by the arrival flow \cite{Whitham} and denoted by $c_i\le 0$. The dissolution speed $c_0<0$ of congested traffic is a characteristic constant with $|c_0| \ge |c_i|$ \cite{NHM,JPhysA}. The average delay time can be determined by averaging over the waiting times in the triangular areas. Note that for the case of an excess greentime ($f_i > u_i$), vehicles may pass the traffic light without any delay.}\label{fIG1}
\end{figure}

Assuming average inflows $A_i$ per lane, the number of vehicles belonging to traffic stream $i$, which must be served within one cycle time $T_{\rm cyc}$, amounts to $A_i T_{\rm cyc}$. In order to avoid the formation of growing queues (i.e. the onset of congestion), the number of served vehicles per cycle time and lane must reach this value. The number of vehicles of traffic stream $i$ potentially served during the green phases $\Delta T_j$ is given by
$\sum_j Q_{ij} \beta_{ij} \Delta T_j$ per lane,  
where $Q_{ij}$ denotes the outflow capacity (discharge flow) per lane, when stream $i$ is served by phase $j$. If we want to have a certain amount of excess capacity to cope with a variability of the inflows, we may demand
\begin{equation}
 (1+\delta_i)A_iT_{\rm cyc} = \sum_j Q_{ij}\beta_{ij} \Delta T_j = \sum_j Q_{ij}\beta_{ij} f_j T_{\rm cyc} 
\label{previous}
\end{equation}
with $\delta_i > 0$. This linear set of equations may be solved for the green time fractions $f_j$. In the following, we will assume the case, where not {\it several} traffic streams $i$ are served in parallel by one and the same green phase, but where each phase $j$ serves a single traffic stream $i$. Then, we may assume $\beta_{ij} = 1$, if $j=i$, and $\beta_{ij}=0$ otherwise.
Furthermore, $Q_{ij} = Q_{ii} = \widehat{Q}_i$. With this, Eq. (\ref{previous}) implies
\begin{equation}
 f_i(u_i,\delta_i) = (1+\delta_i) \frac{A_i}{\widehat{Q}_i} = (1+\delta_i) u_i \, ,
 \label{ef}
\end{equation}
where 
\begin{equation}
u_i = \frac{A_i}{\widehat{Q}_i}
\end{equation} 
is called the {\it utilization} of the service or outflow capacity $\widehat{Q}_i$ (see Fig. \ref{fIG1}). $\delta_i \ge 0$ is a safety factor to cope with variations in the arrival flow (inflow) $A_i$. According to Eq. (\ref{accordi}) we must have
\begin{equation}
0 < 1 - \sum_i f_i = 1 - \sum_i (1+\delta_i) \frac{A_i}{\widehat{Q}_i} \, .
\end{equation}
Otherwise, we will have growing vehicle queues from one signal cycle to the next, corresponding to the congested traffic regime.
\par
Note that, at time $t$, the number $N_i(t)$ of vehicles per lane in the road section reserved for traffic stream $i$
is given by the time integral of the arrival flow $A_i(t)$ minus the departure flow $\gamma_i(t)O_i(t)$:
\begin{equation}
 N_i(t) = \int\limits_{t_{0}}^t dt' \Big[ A_i(t') - \gamma_i(t')
 O_i(t') \Big] \, . 
 \label{contrast}
\end{equation}
Here, the starting time $t_{0}$ must be properly chosen to give the correct initial number of vehicles on road section $i$. During the amber and red time periods, we set the permeability $\gamma_i(t)=0$, as there is no outflow, while $\gamma_i(t)=1$ during green phases. The departure flow $O_i$ per lane is given by the service capacity $\widehat{Q}_i$ per lane, as long as there is a positive number $\Delta N_i(t)>0$ of delayed vehicles. If $Q_{\rm out}$ represents the characteristic outflow from congested traffic per lane into an area of free flow, the {\it overall} service capacity by {\it all} service lanes is given by the minimum of the number of lanes $I_i$ used by vehicle stream $i$ upstream the intersection, and the number $I'_i$ of lanes downstream of it:
\begin{equation}
 I_i \widehat{Q}_i = \min(I_i, I'_i) Q_{\rm out} \, , \quad \mbox{i.e.} \quad 
 \widehat{Q}_i = \min\left(1, \frac{I'_i}{I_i}\right) Q_{\rm out} \, . 
\end{equation}
\par
When the vehicle queue forming behind a traffic light has completely resolved, the outflow from the road section used by vehicle stream $i$ drops from $\widehat{Q}_i$ to a lower value. Then, if the greentime period continues, the outflow $O_i(t)$ per lane corresponds to the arrival flow $A_i(t-{\cal T}_i^0)$ per lane expected at the end of road section $i$ under free flow conditions \cite{JPhysA}. Here, ${\cal T}_i^0 = L_i/V_i^0$ represents the travel time under free flow conditions, which is obtained by division of the length $L_i$ of the road section reserved for stream $i$ by the free speed $V_i^0$. Considering also the above definition of the permeabilities $\gamma_i(t)$ reflecting the time-dependent states of the traffic signal, we have
\begin{equation}
 \gamma_i(t)O_i(t) = \gamma_i(t) \left\{
 \begin{array}{ll}
 \widehat{Q}_i & \mbox{if } \Delta N_i(t) > 0, \\
 A_i(t-{\cal T}_i^0) &\mbox{otherwise.}
 \end{array}\right.
 \label{Oit}
\end{equation}
\par
The number  $\Delta N_i(t)$ of {\it delayed} vehicles per lane at time $t$ on the road section reserved for vehicle stream $i$ can be easily determined as well. In contrast to Eq. (\ref{contrast}), we have to subtract the
integral of the departure flow from the integral of the arrival flow $A_i(t-{\cal T}_i^0)$ expected at the {\it end} of road section $i$ under free flow conditions. Altogether, the number of delayed vehicles can be calculated as
\begin{equation}
 \Delta N_i(t) = \int\limits_{t_{0}}^t dt' \Big[ A_i(t'-{\cal T}_i^0) - \gamma_i(t')O_i(t') \Big] \, .
 \label{grow}
\end{equation}
If the traffic flow is organized as a vehicle platoon and the green phase is synchronized with its arrival at the traffic light, the number of delayed vehicles is zero. However, if the traffic flow $A_i$ is uniform, we find
\begin{equation}
 \Delta N_i(t) = A_i \cdot (t - t_0) - \int\limits_{t_{0}}^t dt' \, \gamma_i(t')O_i(t') \, .  
\end{equation}
Let us assume that $t_0$ denotes the time when the green phase for traffic stream $i$ ended.
Then, the next green phase for this traffic stream starts at time $t'_0 = t_0 + (1-f_i)T_{\rm cyc}$,
as $f_iT_{\rm cyc}$ is the green time period and $(1-f_i)T_{\rm cyc}$ amounts to the sum of the amber and red time periods.  Due to $O_i(t) \ge A_i(t)$ and $O_i(t'_0) > A_i(t'_0)$, $t'_0 - t_0 = (1-f_i)T_{\rm cyc}$ is also the time period after which the maximum number $\Delta N_i^{\rm max}$ of delayed vehicles is reached. 
\par
In case of a uniform arrival of vehicles at the rate $A_i = u_i\widehat{Q}_i$ per lane we have
\begin{eqnarray}
\Delta N_i^{\rm max}(u_i,\{f_j\}) &=& A_i (1-f_i)T_{\rm cyc}(f_i) \nonumber \\
 &=& u_i \widehat{Q}_i (1-f_i)T_{\rm cyc}(\{f_j\})  \, .
\label{dnmax}
\end{eqnarray}
Since $\widehat{Q}_i - A_i$ is the rate at which this vehicle queue can be reduced (considering the further uniform arrival of vehicles at rate $A_i$), it takes a green time period of 
\begin{equation}
T_i(u_i,\{f_j\}) = \frac{A_i(1-f_i)T_{\rm cyc}}{\widehat{Q}_i - A_i} = \frac{u_i(1- f_i 
)}{1-u_i} T_{\rm cyc}(\{f_j\}) \, ,   
\label{Abbau}
\end{equation}
until this vehicle queue is again fully resolved, and newly arriving vehicles can pass the traffic light without any delay.
\par
\begin{figure}[htbp]
\begin{center}
\includegraphics[width=9cm]{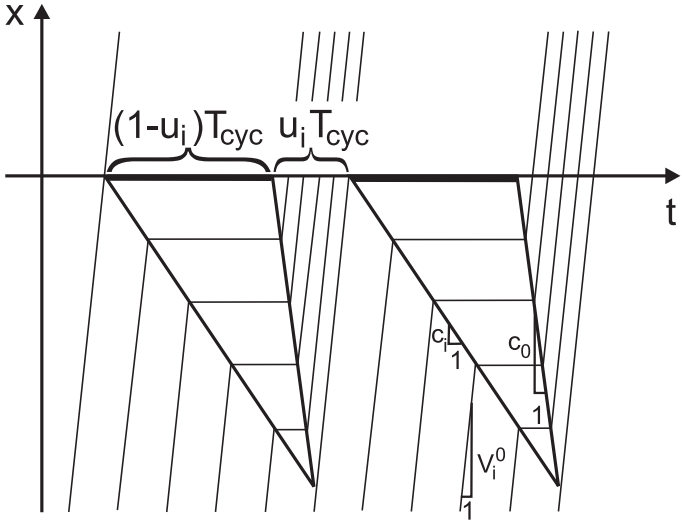}
\end{center}
\caption[]{Schematic illustration of vehicle trajectories and signal operation in the case $f_i=u_i$, where there are no excess green times so that the traffic light is turned red as soon as the vehicle queue has been fully resolved.}
\label{fIG2}
\end{figure}

Finally, let us determine the average delay time of vehicles. If we have a platoon of vehicles which is served by a properly synchronized traffic light, the average delay is ${\cal T}_i^{\rm av} = {\cal T}_i^{\rm min} = 0$. However, if we have a constant arrival flow $A_i$, the average delay ${\cal T}_i^{\rm av}$ of queued vehicles is given by the arithmetic mean $({\cal T}_i^{\rm max}+{\cal T}_i^{\rm min})/2$ of the maximum delay for the first vehicle in the queue behind the traffic light, which corresponds to the amber plus red time period 
\begin{equation}
{\cal T}_i^{\rm max}(\{f_j\}) = (1-f_i)T_{\rm cyc}(\{f_j\}) \, , 
\end{equation} 
and the minimum delay ${\cal T}_i^{\rm min}=0$ of a vehicle 
arriving just at the time when the queue is fully dissolved. To get the average delay, we have to weight this by the percentage of delayed vehicles. While the number of vehicles arriving during the cycle time $T_{\rm cyc}$ is $A_iT_{\rm cyc}$, the number of undelayed vehicles is given by $A_i (\Delta T_i - T_i)$. Considering formulas (\ref{fdef}) and (\ref{Abbau}), the excess green time is
\begin{equation}
\Delta T_i - T_i = f_i T_{\rm cyc} - \frac{u_i(1-f_i)T_{\rm cyc}}{1-u_i} = \frac{f_i - u_i}{1-u_i} T_{\rm cyc} \, . 
\end{equation}
Hence, the percentage of delayed vehicles is
\begin{equation}
\frac{A_i[T_{\rm cyc} - (\Delta T_i - T_i)]}{A_iT_{\rm cyc}} = 1 - \frac{f_i-u_i}{1-u_i} = \frac{1-f_i}{1-u_i} \le 1 \, . \label{perca}
\end{equation}
Altogether, the average delay of {\it all} vehicles is expected to be
\begin{equation}
 {\cal T}_i^{\rm av}(u_i,\{f_j\}) = \frac{(1-f_i)}{(1-u_i)} \, \frac{{\cal T}_i^{\rm max}(\{f_j\})}{2} = \frac{(1-f_i)^2}{1-u_i} \frac{T_{\rm cyc}(\{f_j\})}{2} 
 \label{achtzehn}
\end{equation}
Inserting Eq. (\ref{accordi}) gives
\begin{equation} 
 {\cal T}_i^{\rm av}(u_i,\{f_j\}) = \frac{(1-f_i)^2}{(1-u_i)} \frac{T_{\rm los}}{2(1-\sum_j f_j)}   \, ,
 \label{avtime1}
\end{equation}
while with Eq. (\ref{dnmax}) we obtain
\begin{equation}
  {\cal T}_i^{\rm av}(u_i,\{f_j\}) = \frac{1-f_i}{1-u_i} \frac{\Delta N_i^{\rm max}(u_i,\{f_j\})}{2u_i\widehat{Q}_i} \, . 
  \label{avtime2}
\end{equation}
Therefore, the average delay time is proportional to the maximum queue length $\Delta N_i^{\rm max}$, but the prefactor depends on $f_i = (1+\delta_i)u_i$ (or the safety factor  $\delta_i$, respectively).  In case of no excess green time ($\delta_i = 0$ and $f_i = u_i$), we just have
\begin{eqnarray}
 {\cal T}_i^{\rm av}(\{u_j\})  &=& (1-u_i) \frac{T_{\rm cyc}(\{u_j\})}{2}  = \frac{\Delta N_i^{\rm max}(\{u_j\})}{2A_i} \nonumber\\
 &=& \frac{\Delta N_i^{\rm max}(\{u_j\})}{2u_i\widehat{Q}_i}  
\end{eqnarray}
(see Fig. \ref{fIG2}).
\par
Similarly to the average travel time, we may determine the average queue length. As the average number of delayed vehicles is
$(\Delta N_i^{\rm max}+0)/2$ and a fraction $(1-f_i)/(1-u_i)\le 1$ of vehicles is delayed, together with Eqs. (\ref{dnmax}) and (\ref{achtzehn}) we find
\begin{eqnarray}
\Delta N_i^{\rm av}(u_i,\{f_j\}) &=& \frac{(1-f_i)}{(1-u_i)} \frac{\Delta N_i^{\rm max}(u_i,\{f_j\})}{2} \nonumber \\
&=& u_i \widehat{Q}_i \frac{(1-f_i)^2}{(1-u_i)} \frac{T_{\rm cyc}(\{f_j\})}{2} \nonumber \\
&=& u_i \widehat{Q}_i {\cal T}_i^{\rm av}(u_i,\{f_j\}) 
 \, .\qquad
\label{little}
\end{eqnarray}
This remarkably simple relationship is known in queuing theory as Little's Law  \cite{Little}, which holds for time-averaged variables even in the case of non-uniform arrivals, if the system behaves stable (i.e. the queue length is not {\it systematically} growing or shrinking). It also allows one to establish a direct relationship between the ``average vehicle density'' 
\begin{equation}
\rho_i^{\rm av} = \frac{\Delta N_i^{\rm av}}{L_i}
\label{rhoav1}
\end{equation}
on the road section used by stream $i$ and the utilization $u_i = A_i/\widehat{Q}_i$, namely
\begin{equation}
\rho_i^{\rm av} = \frac{u_i\widehat{Q}_i  {\cal T}_i^{\rm av}}{L_i} \, .  
\label{rhoav2}
\end{equation}
Note that the average density $\rho_i^{\rm av}$ has the same dependencies on other variables as $\Delta N_i^{\rm av}$ or ${\cal T}_i^{\rm av}$, and an additional dependency on $L_i$, i.e. the more natural quantity to use is the queue length $\Delta N_i^{\rm av}$.

\subsection{Efficiency of Traffic Operation}

In reality, the average delay time will depend on the time-dependence of the inflow $A_i(t)$, and on how well the traffic light is coordinated with the arrival of vehicle platoons. In particular, this implies a dependence on the signal offsets. In the best case, the average delay is zero, but in the worst case, it may also be larger than 
\begin{equation}
 \frac{1-u_i}{2} T_{\rm cyc}(\{u_j\}) = \frac{(1-u_i)T_{\rm los}}{2(1-\sum_j u_j)} \, , 
\end{equation}
see Eq. (\ref{achtzehn}) with $f_i = u_i$. We may, therefore, introduce an efficiency coefficient $\epsilon_i$ by the definition
\begin{equation}
{\cal T}_i^{\rm av}(\{u_j\},\epsilon_i)  = (1-\epsilon_i) \frac{(1-u_i) T_{\rm los}}{2(1-\sum_j u_j)}  
\label{defi}
\end{equation}
or, considering Eq. (\ref{little}), equivalently by 
\begin{equation} 
\Delta N_i^{\rm av}(\{u_j\},\epsilon_i)  =  (1-\epsilon_i) u_i \widehat{Q}_i\frac{(1-u_i) T_{\rm los}}{2(1-\sum_j u_j)}  \, .
\label{Nui}
\end{equation}
For $\epsilon_i = 1$, the traffic light is perfectly synchronized with platoons in traffic stream $i$, i.e. vehicles are served without any delay, while for $\epsilon_i = 0$, the delay corresponds to uniform arrivals of vehicles, when no excess green time is given. If the traffic light is not well synchronized with the arrival of vehicle platoons, we may even have $\epsilon_i < 0$. It also makes sense to define an efficiencies not only for the traffic phases, but also for the operation of a traffic light (i.e. the full cycle). This can be done by averaging over all efficiencies $\epsilon_i$ (and potentially weighting them by the number $u_i\widehat{Q}_iT_{\rm cyc}$ of vehicles arriving in one cycle. Therefore, it makes sense to define the intersection efficiency as
\begin{equation}
\epsilon = \frac{\sum_i \epsilon_i u_i\widehat{Q}_i}{\sum_i u_i\widehat{Q}_i} \, .
\end{equation}
Note that, particularly in cases of pulsed rather than uniform arrivals of vehicles, the efficiencies $\epsilon_i$ depend on the cycle time $T_{\rm cyc}$ and, therefore, also on the utilizations $u_i$.
Increasing the efficiency $\epsilon_i$ for one traffic stream $i$ will often (but not generally) reduce the efficiency $\epsilon_j$ of another traffic stream $j$, which poses a great challenge to traffic optimization.
\par
The exact value of the efficiency $\epsilon_i$ depends on many details such as the time-dependence of the arrival flow $A_i(t)$ and its average value $\overline{A}_i$, the length $L_i$ of the road section, and the signal control scheme (fixed cycle time or not, adaptive green phases or not, signal offsets, etc.). These data and the exact signal settings are often not fully available and, therefore, it is reasonable to consider $\epsilon_i$ as fit parameters rather than deriving complicated formulas for them. Nevertheless, we will demonstrate the general dependence on the utilization $u_i$ in the following.
\par
For this, we will study the case of excess green times ($\delta_i > 0$), which are usually chosen to cope with the stochasticity of vehicle arrivals, i.e. the fact that the number of vehicles arriving during one cycle time is usually fluctuating. The choice $\delta_i > 0$, i.e. $f_i > u_i$, also implies 
\begin{equation}
\frac{A_i T_{\rm cyc}}{T_{\rm cyc}} = u_i \widehat{Q}_i < f_i \widehat{Q}_i = \frac{\Delta T_i}{T_{\rm cyc}} \widehat{Q}_i \, . 
\end{equation}
This reflects the well-known observation that the average departure flow $A_i=u_i\widehat{Q}_i$ usually does not reach the value given by the green time fraction $f_i$ times the saturation flow $\widehat{Q}_i$. 
\par
The efficiency $\epsilon_i$ related with a value $\delta_i > 0$ may be derived from Eqs. (\ref{avtime1}) and (\ref{defi}). We obtain
\begin{equation}
		1 - \epsilon_i = \frac{(1-f_i)^2}{(1-u_i)^2} \, \frac{(1-\sum_j u_j)}{(1-\sum_j f_j)} \, ,
		\label{also}
\end{equation}
where $f_i(u_i,\delta_i) = (1+\delta_i)u_i$ according to Eq. (\ref{ef}).
The efficiency $\epsilon_i$ is usually smaller than in the case, where the traffic light is turned red as soon as a vehicle queue has been dissolved (for exceptions see Ref. \cite{slower}). Then, $\epsilon_i < 0$ for $\delta_i > 0$.
Formula (\ref{also}) also allows one to treat the case where the green time fractions $f_i$ and the cycle time $T_{\rm cyc}$ are not adapted to the respective traffic situation, but where a fixed cycle time $T_{\rm cyc} = T_{\rm cyc}^0$ and fixed green time fractions $f_i^0$ are implemented. This corresponds to constant green times 
\begin{equation}
f_i^0T_{\rm cyc}^0(\{f_j^0\}) = \frac{f_i^0 T_{\rm los}}{1 - \sum_j f_j^0} \, .\label{fi0}
\end{equation}
In case of uniform vehicle arrivals, 
we just have to insert the corresponding value $f_i = f_i^0$ into Eq. (\ref{also}) to obtain $\epsilon_i$. 
In the case of non-uniform arrivals, $\epsilon_i$ can be understood as fit parameter of our model, which allows us to adjust our formulas to empirical data and to quantify the efficiency of traffic light operation. In this way, we can also absorb effects of stochastic vehicle arrivals into the efficiency coefficients $\epsilon_i$, which simplifies our treatment a lot.

\section{Fundamental Relationships for Undersaturated Traffic}\label{under}

The travel time is generally given by the sum of the free travel time ${\cal T}_i^0 = L_i/V_i^0$ and the average delay time ${\cal T}_i^{\rm av}$, where $L_i$ denotes the length of the road section used by vehicle stream $i$ and $V_i^0$ the free speed (or speed limit).  With Eq. (\ref{little}), we get
\begin{equation}
{\cal T}_i(\{u_j\},\epsilon_i) = {\cal T}_i^0 + {\cal T}_i^{\rm av}(\{u_j\},\epsilon_i) 
= \frac{L_i}{V_i^0} + \frac{\Delta N_i^{\rm av}(\{u_j\},\epsilon_i)}{u_i\widehat{Q}_i} 
\end{equation}
Inserting Eq. (\ref{Nui}), we can express the travel time solely in terms of the
utilization $u_i$, and we have
\begin{equation}
{\cal T}_i(\{u_j\},\epsilon_i) = \frac{L_i}{V_i^0} + (1-\epsilon_i) \frac{(1-u_i) T_{\rm los}}{2(1-\sum_j u_j)} \, . 
\label{Tiui}
\end{equation}
The formula (\ref{Tiui}) 
constitutes a fundamental relationship between the average travel time ${\cal T}_i^{\rm av}$ on the capacity utilization $u_i$ under the assumptions made (mainly cyclical operation with certain efficiencies $\epsilon_i$). 
Of course, one still needs to specify the factor $(1-\epsilon_i)$. In case of constant arrival rates $A_i$, this factor is given by Eq. (\ref{also}), which finally results in 
\begin{equation}
{\cal T}_i(u_i,\{f_j\}) = \frac{L_i}{V_i^0} + \frac{(1-f_i)^2}{(1-u_i)} \,  \frac{T_{\rm los}}{2(1-\sum_j f_j)} \, .
\end{equation}
After insertion of Eq. (\ref{ef}), we get 
\begin{equation}
{\cal T}_i(\{u_j\},\{\delta_j\}) = \frac{L_i}{V_i^0} + \frac{[1-(1+\delta_i)u_i]^2T_{\rm los}}{(1-u_i)2[1-\sum_j (1+\delta_j)u_j]} \, . 
\label{soso} 
\end{equation}
\par
Sometimes, it is desireable to express the fundamental relationships in terms of the density rather than the utility. Inserting Eq. (\ref{Tiui}) into ${\cal T}_i^{\rm av}(u_i,\epsilon_i) = {\cal T}_i(u_i,\epsilon_i) - L_i/V_i^0$, and this into Eq. (\ref{rhoav2}), we obtain the equation
\begin{equation}
\rho_i^{\rm av}(\{u_j\},\epsilon_i,L_i) = \frac{u_i\widehat{Q}_i}{L_i} (1-\epsilon_i) \frac{(1-u_i) T_{\rm los}}{2(1-\sum_j u_j)} \, , 
\end{equation}
which can be numerically inverted to give the utilization $u_i$ as a function of the scaled densities $\rho_j^{\rm av}L_j/(1-\epsilon_j)$. The calculations are simpler in case of a fixed cycle time $T_{\rm cyc}^0$ and an uniform arrival of vehicles.
By inserting Eq. (\ref{achtzehn}) into (\ref{rhoav2}) 
[and with $f_i = f_i^0$, $T_{\rm cyc} = T_{\rm cyc}^0$, see Eq. (\ref{fi0})], we get 
\begin{eqnarray}
L_i \rho_i^{\rm av} (u_i,\{f_j^0\}) &=& \Delta N_i^{\rm av}(u_i,\{f_j^0\}) \nonumber \\
&=& u_i\widehat{Q}_i  \frac{(1-f_i^0)^2 T_{\rm cyc}^0(\{f_j^0\})}{2(1-u_i)} \, , \label{thirtyeight}
\end{eqnarray}
which finally yields
\begin{equation}
u_i(\rho_i^{\rm av}L_i,\{f_j^0\}) = 
\left( 1 + (1-f_i^0)^2 \frac{\widehat{Q}_iT_{\rm cyc}^0(\{f_j^0\})}{2\rho_i^{\rm av}L_i} \right)^{-1} \, .
\end{equation}
This can be inserted into Eq. (\ref{rhoav2}) to give
\begin{eqnarray}
 {\cal T}_i^{\rm av}(\rho_i^{\rm av}L_i,\{f_j^0\}) &=& \frac{\rho_i^{\rm av}L_i}{u_i(\rho_i^{\rm av}L_i,\{f_j^0\})\widehat{Q}_i} 
 \nonumber \\
 &=& \frac{\rho_i^{\rm av}L_i}{\widehat{Q}_i} + (1-f_i^0)^2 \frac{T_{\rm cyc}^0(\{f_j^0\})}{2} \, . \qquad 
\end{eqnarray}

\subsection{Transition to Congested Traffic}

The utilizations $u_i$ increase proportionally to the arrival flows $A_i$, i.e. they go up during the rush hour. Eventually, 
\begin{equation}
\sum_j f_j = \sum_j (1+\delta_j) u_j \rightarrow 1 \, ,
\end{equation}
which means that the intersection capacity is reached. Sooner or later, there will be no excess capacities anymore, which implies $\delta_i \rightarrow 0$ and $f_i \rightarrow u_i$. In this case, we do not have any finite time periods $\Delta T_i - T_i$, during which there are no delayed vehicles and where the departure flow $O_i(t)$ agrees with the arrival flow $A_i$. Therefore, $O_i(t)= \gamma_i(t)\widehat{Q}_i$ according to Eq. (\ref{Oit}), and certain relationships simplify. For example, the utilization is given as integral of the permeability over one cycle time $T_{\rm cyc}$, divided by the cycle time itself:
\begin{equation}
 u_i = \frac{1}{T_{\rm cyc}} \int\limits_{t_{i0}}^{t_{i0}+T_{\rm cyc}}\!\! dt' \; \gamma_i(t') \, .
 \label{uu}
\end{equation}
Moreover, in the case of constant in- and outflows, i.e. linear increase and decrease of the queue length, 
the average number $\Delta N_i^{\rm av}$ of delayed vehicles is just given by half of the maximum number of delayed vehicles:\footnote{Note that the formulas derived in this paper are exact under the assumption of continuous flows. The fact that vehicle flows consist of discrete vehicles implies deviations from our formulas of upto 1 vehicle, which slightly affects the average travel times as well. In Fig. \ref{fIG2}, for example, the number of vehicles in the queue is $N_i^{\rm max}-1$ rather than $N_i^{\rm max}$. Therefore, the related maximum delay time is $(1-u_i)T_{\rm cyc} (N_i^{\rm max} - 1)/N_i^{\rm max}$, which reduces the average delay time ${\cal T}_i^{\rm av}$ by $(1-u_i)T_{\rm cyc} /(2N_i^{\rm max})$.}
\begin{equation}
 \Delta N_i^{\rm av} = \frac{\Delta N_i^{\rm max}}{2} \, .
\end{equation}
As a consequence, we have
\begin{equation}
{\cal T}_i(\{u_j\}) = \frac{L_i}{V_i^0} + \frac{(1-u_i) T_{\rm los}}{2\big(1-\sum_j u_j\big)}  \, . 
\label{dive}
\end{equation}
The average speed $V_i^{\rm av}$ of 
traffic stream $i$ is often determined by dividing the length $L_i$ of the road section reserved for it by the average travel time ${\cal T}_i = {\cal T}_i^0 + {\cal T}_i^{\rm av}$, which  gives
\begin{equation}
V_i^{\rm av}(\{u_j\}) = \frac{L_i}{{\cal T}_i(\{u_j\})}  = \left( \frac{1}{V_i^0} + \frac{(1-u_i) T_{\rm los}}{2L_i(1-\sum_j u_j)} \right)^{-1} \, , 
 \label{remembe}
\end{equation}
and can be generalized with Eq. (\ref{defi}) to cases with an efficiencies $\epsilon_i \ne 0$:
\begin{equation}
V_i^{\rm av}(\{u_j\},\epsilon_i) = \left( \frac{1}{V_i^0} + (1-\epsilon_i)\frac{(1-u_i) T_{\rm los}}{2L_i(1-\sum_j u_j)} \right)^{-1} \, . 
\label{Viui}
\end{equation}
Note, however, that the above formulas for the average speed are implicitly based on a harmonic rather than an arithmetic average. When correcting for this, Eq. (\ref{remembe}), for example, becomes
\begin{equation}
	V_i^{\rm av}(\{u_j\}) = \frac{L_iu_i\widehat{Q}_i}{\Delta N_i^{\rm max}(\{u_j\})} \ln \left| 1 + \frac{\Delta N_i^{\rm max}(\{u_j\})}{u_i\widehat{Q}_i{\cal T}_i^0}\right| \, .  
\end{equation}
As is shown in Appendix \ref{ap1}, this has a similar Taylor approximation as the harmonic average (\ref{Viui}). The latter is therefore reasonable to use, and it is simpler to calculate.
\par
As expected from queuing theory, the average travel time (\ref{dive})
diverges, when the sum of utilizations reaches the intersection capacity, i.e. $\sum_j u_j \rightarrow 1$.
In this practically relevant case, the traffic light would not switch anymore, which would frustrate drivers. For this reason, the cycle time is limited to a finite value 
\begin{equation}
T_{\rm cyc}^{\rm max}(\{u_j^0\}) = \frac{T_{\rm los}}{1-\sum_j u_j^0} \, , 
\end{equation}
where typically $u_j^0 \le u_j$. 
This implies that the sum of utilizations must fulfill
\begin{equation} 
 \sum_j u_j \le \sum_j u_j^0 = 1 - \frac{T_{\rm los}}{T_{\rm cyc}^{\rm max}} \, . 
\end{equation}
As soon as this condition is violated, we will have an increase of the average number of delayed vehicles in time, which characterizes the congested regime discussed in the next section.

\section{Fundamental Relationships for Congested Traffic Conditions}\label{congest}

In the congested regime, the number of delayed vehicles does not reach zero anymore, and platoons cannot be served without delay. Vehicles will usually have to wait several cycle times until they can finally pass the traffic light. This increases the average delay time enormously. It also implies that there are no excess green times, which means $\delta_i = 0$. Consequently, we can also assume $O_i(t) = \widehat{Q}_i$, as long as the outflow from road sections during the green phase is not (yet) obstructed (otherwise see Sec. \ref{over}). Formula (\ref{grow}) applies again and implies for time-independent arrival flows $A_i$ 
\begin{eqnarray}
 \Delta N_i(t_{i0}+kT_{\rm cyc}^{\rm max})&=&  \Delta N_i(t_{i0}) + \!\!\!\!\int\limits_{t_{i0}}^{t_{i0}+kT_{\rm cyc}^{\rm max}}\!\!\!\! dt' \Big[ A_i - \gamma_i(t')\widehat{Q}_i \Big] \nonumber \\
 &=&   \Delta N_i(t_{i0})  + (A_i - u_i^0\widehat{Q}_i)kT_{\rm cyc}^{\rm max} \, ,
\end{eqnarray}
where 
\begin{equation}
u_i^0 = \frac{1}{kT_{\rm cyc}^{\rm max}} \!\!\!\int\limits_{t_{i0}}^{t_{i0}+kT_{\rm cyc}^{\rm max}}\!\!\!\! dt' \; \gamma_i(t') < u_i = \frac{A_i}{\widehat{Q}_i}
\end{equation}
is the green time fraction of the cycle time $T_{\rm cyc}^{\rm max}$ reserved for vehicle stream $i$. Therefore, the number of delayed vehicles grows by an amount $(A_i - u_i^0\widehat{Q}_i)T_{\rm cyc}^{\rm max} = (u_i - u_i^0)\widehat{Q}_ikT_{\rm cyc}^{\rm max} $ in each cycle time $T_{\rm cyc}^{\rm max}$. The minimum number during one cycle is
\begin{equation}
 \Delta N_i^{\rm min} (u_i,k) = \Delta N_i(t_{i0}) + (u_i - u_i^0)\widehat{Q}_ikT_{\rm cyc}^{\rm max} \, ,
 \label{Nmin}
\end{equation}
and considering Eq. (\ref{dnmax}), the maximum number of delayed vehicles is
\begin{equation}
\Delta N_i^{\rm max}(u_i,k) = \Delta N_i^{\rm min} (u_i,k) + u_i (1-u_i^0)\widehat{Q}_iT_{\rm cyc}^{\rm max} \, .
\label{Nm}
\end{equation}
Because of $\Delta N_i^{\rm av} = (\Delta N_i^{\rm min} +\Delta N_i^{\rm max} )/2$ and $A_i = u_i \widehat{Q}_i$,  the average number of delayed vehicles is
\begin{eqnarray}
\Delta N_i^{\rm av}(u_i,\{u_j^0\},k) &=& \bigg( (u_i - u_i^0) k \nonumber \\
& & + \frac{u_i (1-u_i^0)}{2} \bigg) \widehat{Q}_iT_{\rm cyc}^{\rm max}(\{u_j^0\})\, . \qquad
\label{trila}
\end{eqnarray}
Obviously, the average density 
\begin{eqnarray}
\rho_i^{\rm av} (u_i,\{u_j^0\},L_i,k) &=& \bigg( (u_i - u_i^0) k \nonumber \\
& & + \frac{u_i (1-u_i^0)}{2} \bigg) \frac{\widehat{Q}_iT_{\rm cyc}^{\rm max}(\{u_j^0\})}{L_i} \qquad 
\end{eqnarray}
is obtained by dividing the previous formula by $L_i$. 
\par
\begin{figure}[htbp]
\begin{center}
\includegraphics[width=9cm]{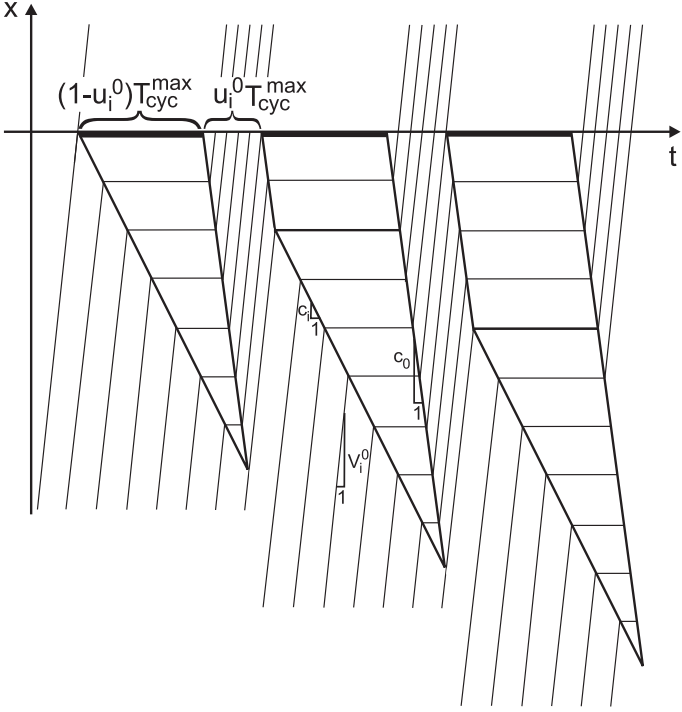}
\end{center}
\caption[]{Schematic illustration of growing vehicle queues in the congested regime. Since not all vehicles can be served within one green time $u_i^0T_{\rm cyc}^{\rm max}$, vehicles are forced to make additional stops, which correspond to the rhomboidal parts of the queue in the above time-space diagram. The triangular part, in contrast, represents newly arriving vehicles.}
\label{fIG3}
\end{figure}

It seems logical to use Eq. (\ref{trila})  to determine the average delay time as 
\begin{eqnarray}
{\cal T}_i^{\rm av}(u_i,\{u_j^0\},k) &=& \frac{\Delta N_i^{\rm av}(u_i,\{u_j^0\},k)}{A_i} \nonumber \\
&=& \left(1 - \frac{u_i^0}{u_i}\right) kT_{\rm cyc}^{\rm max} (\{u_j^0\}) \nonumber \\
&+& \frac{(1-u_i^0)T_{\rm cyc}^{\rm max}(\{u_j^0\})}{2} \, . 
\end{eqnarray}
However, this turns out to be {\it not} correct, as some of this time is actually not lost, but used to move forward (see Fig. \ref{fIG3}).
After this first simple analysis, we need to look at the problem more carefully, and perform the calculation without the use of Eq. (\ref{little}), as Little's Law is not applicable in case of non-stable, systematically growing queues. 
\par
Again, we apply the fact that the number of vehicles arriving in one cycle is $A_iT_{\rm cyc}^{\rm max}$, while the number of vehicles served during one green time period of duration $u_i^0T_{\rm cyc}^{\rm max}$ is given by $\widehat{Q}_i u_i^0T_{\rm cyc}^{\rm max}$. The difference of both numbers is added to the growing vehicle queue. The quotient of the number  $kA_iT_{\rm cyc}^{\rm max}$ of vehicles arriving in $k$ cycles and the 
number $\widehat{Q}_i u_i^0T_{\rm cyc}^{\rm max}$ of vehicles served during one green time period, when rounded down, corresponds to the number $n_{\rm s}$ of additional stops needed by newly arriving vehicles. Therefore,  with $A_i = u_i \widehat{Q}_i$ we have
\begin{equation}
n_{\rm s}(u_i,(\{u_j^0\}),k) = \left\lfloor \frac{A_ikT_{\rm cyc}^{\rm max}(\{u_j^0\})}{\widehat{Q}_i u_i^0T_{\rm cyc}^{\rm max}(\{u_j^0\})} \right\rfloor
= \left\lfloor \frac{u_ik}{u_i^0} \right\rfloor \, .\label{this}
\end{equation}
Here, $\lfloor x \rfloor$ represents the floor function rounding to lower integers (i.e. cutting the digits after the period, if $x>0$). Assuming that $t_{i0}$ is the time at which congestion sets in, Eq. (\ref{this}) can be generalized to continuous time $t$, allowing us to estimate the number of additional stops of a vehicle arriving at time $t$:
\begin{equation}
n_{\rm s}(u_i,\{u_j^0\},t) = \left\lfloor \frac{A_i(t-t_{i0})}{\widehat{Q}_i u_i^0T_{\rm cyc}^{\rm max}(\{u_j^0\})} \right\rfloor
= \left\lfloor \frac{u_i(t-t_{i0})}{u_i^0T_{\rm cyc}^{\rm max}(\{u_j^0\})} \right\rfloor \, .
\label{ns}
\end{equation}
\par 
As Fig. \ref{fIG3} shows, the delay time by each of the $n_{\rm s}$ additional stops is $(1-u_i^0)T_{\rm cyc}^{\rm max}$. 
Moreover, we can see that the triangular part of the vehicle queue in the space-time plot gives a further contribution to the delay time
of vehicles. Applying Eq. (\ref{achtzehn}) with $f_i = u_i = u_i^0$, 
the average time delay in this triangular part is $(1-u_i^0)T_{\rm cyc}^{\rm max}/2$,
i.e. the arithmetic average between zero and the sum of the red and amber time period (amounting to $(1-u_i^0)T_{\rm cyc}^{\rm max}$). In summary,  the  delay time of a newly arriving vehicle at time $t$ (when averaging over the triangular part for the sake of simplicity), is
\begin{eqnarray}
{\cal T}_i^{\rm av}(u_i,t) &=& (1-u_i^0)\frac{T_{\rm cyc}^{\rm max}}{2} + \left\lfloor \frac{u_i(t-t_{i0})}{u_i^0T_{\rm cyc}^{\rm max}} \right\rfloor (1-u_i^0)T_{\rm cyc}^{\rm max} \nonumber \\
&=& \left( \frac{1}{2} + \left\lfloor \frac{u_i(t-t_{i0})}{u_i^0T_{\rm cyc}^{\rm max}} \right\rfloor \right) (1-u_i^0)T_{\rm cyc}^{\rm max}  \, . \label{Tiav1}
\end{eqnarray}
Accordingly, the average travel time does not only grow with time $t$ (or the number $k$ of cycles passed), it also grows {\it stepwise} due to the floor function $\lfloor x \rfloor$.  
\par
When we also average formula (\ref{Tiav1}) over its steps, we obtain the approximate relationship
\begin{eqnarray}
{\cal T}_i^{\rm av}(u_i,t) &\approx& \left(  \frac{u_i(t-t_{i0})}{u_i^0T_{\rm cyc}^{\rm max}} \right) (1-u_i^0)T_{\rm cyc}^{\rm max} \nonumber \\
&=&  u_i(t-t_{i0}) \frac{(1-u_i^0)}{u_i^0} \, ,
\end{eqnarray}
where it is important to consider that the floor function $\lfloor x \rfloor$ is shifted by $0.5$ with respect to the function $\mbox{round}(x)$, which rounds to the closest integer: $\mbox{round}(x) = \lfloor x +0.5 \rfloor$.
Moreover, when averaging over the last ($k+1$st) cycle we get
\begin{equation}
{\cal T}_i^{\rm av}(u_i,k) \approx u_i \left(k + \frac{1}{2} \right) \frac{1-u_i^0}{u_i^0} T_{\rm cyc}^{\rm max} \, .
\label{kaa}
\end{equation}
It is also possible to replace the dependence on the number $k$ of cycles by a dependence on the average density of delayed vehicles by applying Eq. (\ref{trila}). In this way, we obtain 
\begin{equation}
k+\frac{1}{2} = \frac{\Delta N_i^{\rm av}(u_i,\{u_j^0\},k)}{(u_i-u_i^0)\widehat{Q}_iT_{\rm cyc}^{\rm max}(\{u_j^0\})} - \frac{u_i^0(1-u_i)}{2(u_i - u_i^0)} \, . 
\end{equation}
Therefore, while in Sec. \ref{under} we could express the average travel time and the average velocity either in dependence of
the average density $\rho_i^{\rm av}$ or the utilization $u_i$ alone, we now have a dependence on both quantities.
\par
Finally note that Eqs. (\ref{Tiav1}) to (\ref{kaa}) may be generalized to the case where the arrival rate of vehicles is not time-independent. This changes the triangular part of Fig. \ref{fIG3}. In order to reflect this, the corresponding contribution $(1-u_i^0)T_{\rm cyc}^{\rm max}/2$ may, again, be multiplied with a prefactor $(1-\epsilon_i)$, which defines an efficiency $\epsilon_i$. While $\epsilon_i=0$ corresponds to the previously discussed case of an uniform arrival of vehicles, $\epsilon_i = 1$ reflects the case where a densely packed platoon of vehicles arrives at the moment when the last vehicle in the queue has started to move forward.  

\section{Fundamental Relationships for Oversaturated Traffic Conditions}\label{over}

\begin{figure}[htbp]
\begin{center}
\includegraphics[width=9cm]{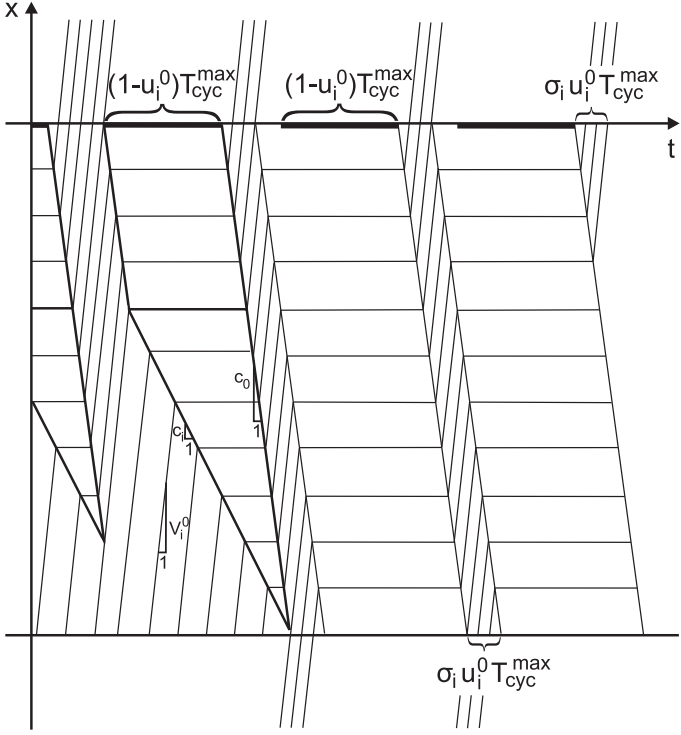}
\end{center}
\caption[]{Schematic illustration of the service of vehicle queues, when the road section is fully congested. The lower horizontal line indicates the location of the upstream end of the road section. It can be seen that vehicles are stopped several times, and that new vehicles can only enter when some vehicles have been served by the traffic light located at the $t$-axis, and the  space freed up by this  has reached the end of the vehicle queue.  For similar considerations see Ref. \cite{Eichler}. Note that the characteristic speed $c_0$ of the shock fronts (which corresponds to the slope of the congested flow-density relationship for a road section without the consideration of traffic lights) is different from the slope of the {\it urban} fundamental diagram, because of the effect of signal offsets and delays \cite{DagGer}. Therefore, $c_0$ should be understood as fit parameter, here.} 
\label{fIG3a}
\end{figure}

We have seen that, under congested conditions, the number of delayed vehicles is growing on average. Hence, the vehicle queue will eventually fill the road section reserved for vehicle stream $i$ completely. Its maximum storage capacity per lane for delayed vehicles is
\begin{equation}
\Delta N_i^{\rm jam}(L_i) = L_i \rho_i^{\rm jam} \, , 
\end{equation}
where $\rho_i^{\rm jam}$ denotes the maximum density of vehicles per lane. The road section becomes completely congested at the time 
\begin{equation}
t = t_{i0}+kT_{\rm cyc}^{\rm max} + \Delta t \, ,
\end{equation}
when $\Delta N_i^{\rm min}(u_i,k) +A_i \Delta t$ according to Eq. (\ref{Nmin}) reaches the value $\Delta N_i^{\rm jam}$, which implies
\begin{equation}
 \Delta t = \frac{\Delta N_i^{\rm jam} - \Delta N_i^{\rm min}(u_i,k)}{A_i} \, , 
\end{equation}
where 
\begin{equation}
 k = \left\lfloor \frac{\Delta N_i^{\rm jam}}{(u_i - u_i^0)T_{\rm cyc}^{\rm max}} \right\rfloor \, , 
\end{equation}
see Eq. (\ref{Nmin}).
The number $n_{\rm s}+1$ of stops is given by Eq. (\ref{ns}). See Fig. \ref{fIG3a} for an illustration.
\par
Complete congestion causes spillover effects and obstructs the arrival of upstream vehicles, even though the green phase would, in principle, allow them to depart. When these spill-over effects set in, we are in the over-saturated regime, 
and only a certain fraction $\sigma_i$ of the green phase of duration $u_i^0 T_{\rm cyc}^{\rm max}$ can be used, where $0 \le \sigma_i \le 1$. Note that $\sigma_i$ decreases in time as the number of blocked subsequent road sections grows. This may eventually lead to gridlock in a large area of the urban traffic system.
\par
In the case $\sigma_i < 1$,  $f_i=u_i^0$ must essentially be replaced by $f_i=\sigma_iu_i^0$ in the fundamental relationships for congested traffic, and the lost service time becomes 
\begin{equation}
T_{\rm los} = \sum_i \Big[ \tau_i + (1-\sigma_i) u_i^0 T_{\rm cyc}^{\rm max} \Big] \, .
\end{equation}
That is, the reduced service time may be imagined like an extension of the amber time periods. Therefore, it would make sense to {\it reduce} the cycle time to a value $T_{\rm cyc} < T_{\rm cyc}^{\rm max}$ 
in the oversaturated regime. 
\par
Note, however, that the travel times on the road section reserved for stream $i$ are not growing anymore, because the road section is limited to a length $L_i$. This allows us to determine the corresponding travel time on the link as follows: 
The number of vehicles served per cycle time is $\sigma_i u_i^0 T_{\rm cyc}^{\rm max}\widehat{Q}_i$. For this reason, the average travel time can be estimated as
\begin{equation}
{\cal T}_i(u_i,\sigma_i,L_i) = \frac{\Delta N_i^{\rm jam}(L_i)}{\sigma_i u_i^0 T_{\rm cyc}^{\rm max}} T_{\rm cyc}^{\rm max}
= \frac{L_i \rho_i^{\rm jam}}{\sigma_i u_i^0 \widehat{Q}_i} \, ,
\end{equation}
and the average delay time ${\cal T}_i^{\rm av} = {\cal T}_i - {\cal T}_i^0$ as
\begin{equation}
{\cal T}_i^{\rm av}(u_i,\sigma_i,L_i) 
= \frac{L_i \rho_i^{\rm jam}}{\sigma_i u_i^0 \widehat{Q}_i} - \frac{L_i}{V_i^0} \, .
\label{Tiover}
\end{equation}
Note that these values are now {\it in}dependent of both, the utilization and the average density, as soon as the latter assumes the value $\rho_i^{\rm av} = \rho_i^{\rm jam}$, corresponding to a fully congested road section.

\subsection{Transition from Oversaturated to Undersaturated Traffic Conditions}

If the arrival flow $A_i$ after the rush hour drops below the value of $\sigma_i u_i^0 T_{\rm cyc}^{\rm max}$, the vehicle queue will eventually shrink, and the road section used by vehicle stream $i$ enters from the oversaturated into the congested regime. The formulas for the evolution of the number of delayed vehicles are analogous to Eqs. (\ref{Nmin}) and (\ref{Nm}). The queue length starts with $\Delta N_i^{\rm jam}$ and is reduced by 
$(u_i - \sigma_i u_i^0)\widehat{Q}_ikT_{\rm cyc} 
<0$ in each cycle of length $T_{\rm cyc}$. 
Counting the number of cycles since the re-entering into the congested regime by $k'$, we have
\begin{equation}
 \Delta N_i^{\rm min} (u_i,k') = \Delta N_i^{\rm jam} + (u_i - \sigma_i u_i^0)\widehat{Q}_ik'T_{\rm cyc} \, ,
 \label{Nmink}
\end{equation}
and considering Eq. (\ref{dnmax}), the maximum number of delayed vehicles is
\begin{equation}
\Delta N_i^{\rm max}(u_i,k') = \Delta N_i^{\rm min} (u_i,k') 
+ u_i (1-\sigma_i u_i^0)\widehat{Q}_iT_{\rm cyc} \, . 
\end{equation}
As soon as $\Delta N_i^{\rm min}(u_i,k')$ reaches zero, the road section used by vehicle stream $i$ enters the undersaturated regime. Before, the number of stops of vehicles joining the end of the vehicle queue
are expected to experience an number $n_{\rm s}+1$ of stops with
\begin{equation}
n_{\rm s}(k') 
= \left\lfloor \frac{u_ik'}{\sigma_iu_i^0} \right\rfloor \, ,
\end{equation}
compare Eq. (\ref{this}).

\section{Summary and Outlook}\label{sum}

\begin{figure}[htbp]
\begin{center}
\includegraphics[height=9cm,angle=-90]{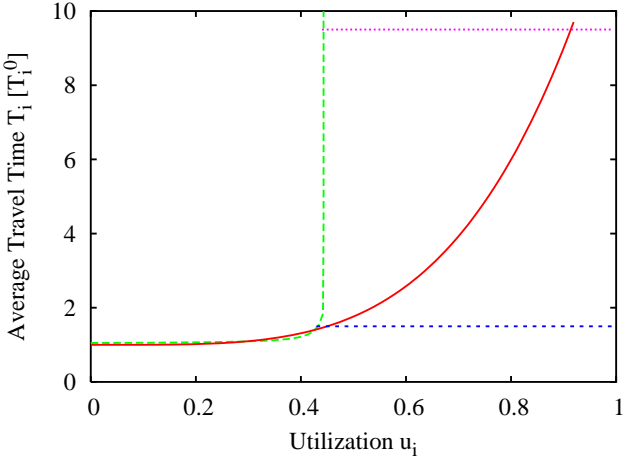}
\end{center}
\caption[]{Schematic illustration of the capacity restraint function ({\protect\ref{capres}}) 
that the Bureau of Public Roads recommends to use {\protect\cite{BPR}} 
(solid line), together with analytical results of this paper (dashed lines). Travel time ${\cal T}_i$ is measured in units of ${\cal T}_i^0$. The green long-dashed line shows that the travel time diverges at $u_i \approx 0.45$, if two green phases during one cycle and identical utilizations $u_1 = u_2$ of both associated road sections are assumed, furthermore, if the parameters are set to $T_{\rm los} = 0.1 {\cal T}_i^0$ 
and $\delta = 0.1$. (A generalization to traffic operation with more than two green phases is easily possible.) When the cycle time is limited to a finite value ${\cal T}_{\rm cyc}^{\rm max}$ to avoid infinite delay times in one of the vehicle queues, one will have growing vehicle queues and increasing travel times in both road sections. Therefore, the travel time can assume any value above the lower dashed horizontal line. The link travel time is only limited by the above dashed horizontal line, which corresponds to the situation where the vehicle queue fills the road section completely. Note that the capacity restraint function (solid line) averages over all travel time measurements for a given utilization $u_i$. In the right part of the diagram this concerns measurements that depend on the duration of congestion and scatter between both horizontal dashed lines. Here, the curve is shown for $\alpha_i = 0.5$, $\beta_i=4$, and $A_i/C_i = u_i/0.45$.}
\label{fIG4}
\end{figure}

Based on a few elementary assumptions, we were able to derive fundamental relationships for the average travel time ${\cal T}_i^{\rm av}$ and average velocity $V_i^{\rm av}$. These relationships are functions of the utilization $u_i$ of the service capacity of a cyclically signalized intersection and/or the average number $\Delta N_i^{\rm av}$ of delayed vehicles (or the average density $\rho_i^{\rm av}$ of vehicles in the road section of length $L_i$ reserved for traffic stream $i$). We found different formulas, (1) for the undersaturated regime,
(2) for the congested regime, and (3) for the oversaturated regime. While we also discussed situations, where fixed cycle times $T_{\rm cyc}^0$ are applied, we primarily focussed on situations, where the cycle time is adjusted to the utilization $u_i$ (in the undersaturated regime) and to the effectively usable green time fraction $\sigma_i$ (in the oversaturated regime). Our results are summarized in Fig. \ref{fIG4}), where also a comparison with the capacity restraint function (\ref{capres}) is made.
\par
The formulas for the {\it non-congested regime} can be either expressed as non-trivial functions of the utilization $u_i$ or the average queue length $\Delta N_i^{\rm av}$ (or the average density $\rho_i^{\rm max}$). They contain a fit parameter $\epsilon_i$, which reflects effects of variations in the arrival flow and relates to the efficiency of traffic signal operation in terms of synchronizing with vehicle platoons. In the best case, delay times are zero, which shows the great optimization potential for traffic control in this regime.
\par
In the {\it congested regime}, the number of delayed vehicles grows in time, and the majority of vehicles is stopped several times by the same traffic light. Therefore, the average travel time does not only depend on the utilization $u_i$, but also on the average vehicle queue $\Delta N_i^{\rm av}$ (or the average density $\rho_i^{\rm av}$). Although the traffic light control can still improve the average travel times by synchronizing with the arrival of vehicles, the related efficiency effect is rather limited.  
\par
In the {\it oversaturated regime}, the storage capacity of the road section is fully occupied by delayed vehicles, which obstructs the arrival flow. 
Therefore,  the average travel time of a road section reaches a constant maximum value. Nevertheless, the travel time of {\it vehicles} increases further in time due to spillover effects, which trigger the spreading of congestion to upstream road sections. The actually usable fraction of the green time period is described by a parameter $\sigma_i$. Synchronization can still reach some improvements. The most favorable control is oriented at a fluent upstream propagation of the little remaining free space (the difference between $\Delta N_i^{\rm max}$ and $\Delta N_i^{\rm min}$). That is, rather than minimizing the delay of downstream moving vehicle platoons, one should now minimize the delay in filling upstream moving gaps \cite{movinggaps}. Furthermore, note that the free travel time ${\cal T}_i^0 = L_i /V_i^0$ and the delay time in the oversaturated regime are proportional to the length $L_i$ of the road section, while the delay times in the undersaturated and congested regimes are independent of $L_i$.
\par
Based on the above results, it is obvious that it cannot be very successful to describe the travel time of a link by a capacity restraint function which depends on $A_i/C_i = u_i/u_i^0$ only. This basically means an averaging over data that, in principle, are also dependent on the average queue length $\Delta N_i^{\rm av}$  (or on the time passed since the onset of congestion). It is, therefore, no wonder that empirical data of travel times as a function of the utilization scatter so enormously  in the congested and oversaturated regimes (see e.g. Ref. \cite{Tu}), that a fitting of the data to functional dependencies of any kind does not make much sense. 
\par
This has serious implications for transport modeling, as capacity restraint functions such as formula (\ref{capres}) are used for modeling route choice and, hence, for traffic assignment. Based on the necessary revision of this formula and comparable ones, all traffic scenarios based on these formulas should be critically questioned. Given the computer power of today, it would not constitute a problem to perform a dynamic traffic assignment and routing based on the more differentiated formulas presented in this paper.

\subsection{Transferring the Link-Based Urban Fundamental Diagrams to an Area-Based One}

We may finally ask ourselves, whether the above formulas would also allow one to make predictions about the average travel times and speeds for a whole {\it area} of an urban traffic network, rather than for single road sections (``links'') only. This would correspond to averaging over the link-based fundamental diagrams of that area. For the sake of simplicity, let us assume for a moment that the parameters of all links would be the same, and derive a velocity-density diagram from the
relationship (\ref{Taking}) between average vehicle speed $V_i^{\rm av}$ and the capacity utilization $u_i$. Taking into account $V_i^0 = L_i/{\cal T}_i^0$ and dropping the index $i$, we can write
\begin{equation}
V^{\rm av}(u) = \frac{V^0\ln \big\{ 1 + [1-f(u)]T_{\rm cyc}/{\cal T}^0 \big\}}{(1-u)T_{\rm cyc}/{\cal T}^0}
+ V^0 \frac{f(u)-u}{1-u} \, . 
\label{FI1}
\end{equation}
with $f(u) = (1+\delta)u$. The cycle time
\begin{equation}
T_{\rm cyc} = \frac{T_{\rm los}}{1 - sf(u)} \label{FI2} 
\end{equation}
follows from Eq. (\ref{accordi}), assuming $s$ signal phases with $f_i = f$ for simplicity. Note that the previous formulas for $\rho_i^{\rm av}$ denote the average density of {\it delayed} vehicles, while the average density of all vehicles (i.e. delayed and freely moving ones) is given by
\begin{equation}
\rho(u) = \frac{N(u)}{L} = \frac{N(u)/{\cal T}(u)}{L/{\cal T}(u)} = \frac{A(u)}{V^{\rm av}(u)} 
= \frac{u\,\widehat{Q}}{V^{\rm av}(u)} \, , 
\label{Lirho}
\end{equation}
where $N = A{\cal T} = u\widehat{Q}{\cal T}$ denotes the average number of vehicles on a road section of length $L$. Plotting $V^{\rm av}(u)$ over $\rho(u)$ finally gives a speed-density relationship $V^{\rm av}(\rho)$. 
\par
We have now to address the question of what happens, if we average over the different road sections of an urban area. Considering the heterogeneity of the link lengths $L_i$, efficiencies $\epsilon_i$, utilizations $u_i$, and densities $\rho_i^{\rm av}$, one could think that the spread in the data would be enormous. However, a considerable amount of smoothing results from the fact that in- and outflows of links within the studied urban area cancel out each other, and it does not matter whether a vehicle is delayed in a particular link, or in the previous or subsequent one. Therefore, the resulting fundamental diagrams for urban areas are surprisingly smooth \cite{Geroliminis,Simdata}. The details of the curves, however, are expected to depend not only on the average density, but also on the density distribution, the signal operation schemes, and potentially other factors as well. 
\par
When averaging over different links, we have to study the effects of the averaging procedure on the density and the speed. The density just averages linearly: Thanks to Little's law \cite{Little}, the formula (\ref{Lirho}) can also be applied to an urban area, as long as the average number of vehicles in it is stationary. However, as the link-based fundamental diagram between the flow $Q(\rho) = \rho V^{\rm av}(\rho)$ and the density $\rho$ is convex, evaluating the flow at some average density overestimates the average flow.\footnote{When averaging over speed values, they have to be weighted by the number of vehicles concerned, i.e. by the density. This comes down to determining an arithmetic average of the flow values and dividing the result by the arithmetic average of the related densities. Hence, an overestimation of the average flow also implies an overestimation of the average velocity. However, considering the curvature of $Q(\rho)$ and knowing the variability of the density $\rho$ allows one to estimate correction terms.} This also implies that, when the relationship $V^{\rm av}(\rho)$ is transferred from single links to urban areas, the average speed is overestimated for a given density, as Ref. \cite{DagGer} has shown. 
\par\begin{figure}[htbp]
\begin{center}
\includegraphics[width=8.8cm]{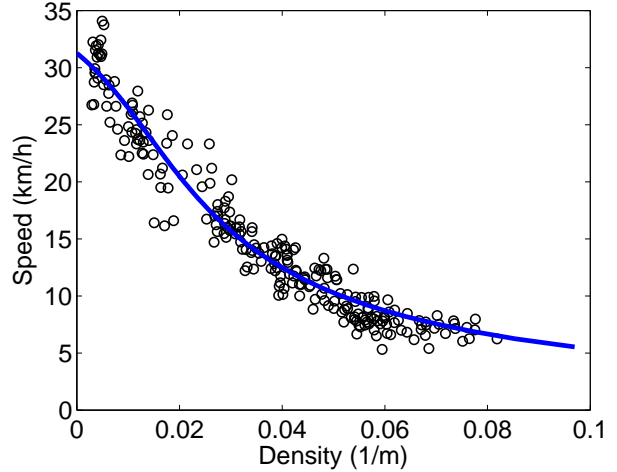}
\end{center}
\caption[]{Fundamental velocity-density relationship for a central area of Yokohama. Small circles correspond to empirical data by Kuwahara as evaluated by Daganzo and Geroliminis \cite{DagGer}. The fit curve corresponds to the theoretically derived equations (\ref{FI1}) to (\ref{Lirho}), where the outflow (discharge flow) $\widehat{Q} = 1800$~veh./hour/lane and the free speed $V^0 = 50$~km/h have been fixed. The only fit parameters were $\delta = 0.1$, $T_{\rm los}/{\cal T}_0 = 1.4$, and $s=3$.\footnotemark[3]} 
\label{sechs}
\end{figure}\footnotetext[3]{This corresponds to 10\% extra green time, an average distance between successive traffic lights of roughly 100 meters (depending on $T_{\rm los}$), and an average of 3 traffic phases (which appears plausible, considering that there are many one-way roads, which need less than 4 phases in one cycle time). Consequently, all parameters are quite reasonable (see also Ref. \cite{DagGer}). Note that effects of oversaturation did not have to be considered in Fig. \ref{sechs}. This is, in fact, consistent with pictures from Google Earth.}
Despite this expected deviation in heterogeneous and inhomogeneously used road networks,
and despite the many other simplifications, the curve $V^{\rm av}(\rho)$ fits empirical data of the speed-density relation in an urban area quite well. Figure \ref{sechs} displays empirical data obtained for the center of Yokohama \cite{Geroliminis} together with a fit of the theoretical speed-density relationship $V^{\rm av}(\rho)$, where only the three parameters $V_0$, $\delta$, and $V^* = L/T_{\rm los}$ were adjusted. Surprisingly, the effects of network interactions could be sufficiently well represented by a single parameter $\delta$, which relates to the efficiency $\epsilon$ of road sections according to Eq. (\ref{also}). This approximation seems to work in situations close enough to a statistical equilibrium (when the number of vehicles in the urban area does not change too quickly). 
\par
In contrast, for an understanding of the {\it spreading} dynamics of congestion patterns, we expect that one must study the interaction between the flow dynamics and the network structure. This difficult subject goes beyond the scope of this paper and beyond what is doable at the moment, but it will be interesting to address it in the future. 

\begin{acknowledgement}
The author thanks for an inspiring presentation by  Carlos Daganzo, for useful comments by Stefan L\"ammer, and intresting discussions with Nikolas Geroliminis, who was also kind enough to provide the empirical data from the center of Yokohama displayed in Fig.~6, see Fig. 7 in Ref. \cite{Geroliminis}. He extracted these from original data of GPS-equipped taxis by Prof. Masao Kuwahara from the University of Tokyo. The fit of the theoretically predicted relationship to the empirical data was carried out by Anders Johansson. Furthermore, the author is grateful for partial support by the Daimler-Benz Foundation Project 25-01.1/07 on BioLogistics, the VW Foundation Project I/82 697, the NAP project KCKHA005 ``Complex Self-Organizing Networks of Interacting Machines: Principles of Design, Control, and Functional Optimization'', and the ETH Competence Center 'Coping with Crises in Complex Socio-Economic  Systems' (CCSS) through ETH Research Grant CH1-01-08-2.

\end{acknowledgement}

\appendix
\section{Determination of Average Travel Times and Velocities} \label{ap1}
Let $f(x)$ be a function and $w(x)$ a weight function. Then, the average of the function between $x=x_0$ and $x=x_1$ is defined as
\begin{equation}
 \frac{\int\limits_{x_0}^{x_1} dx' \; w(x')f(x')}{ \int\limits_{x_0}^{x_1} dx' \; w(x')} \, .
\end{equation}
In case of uniform arrivals of vehicles, we have a functional relationship of the form $f(x) = a + bx$ for the travel time, and the weigth function is constant, i.e. $w(x) = w$. Here, $a={\cal T}_i^0$, $x = \Delta N_i$, and $b =1/A_i = 1/(u_i\widehat{Q}_i)$.  With $x_0=0$ and $x_1 = \Delta N_i^{\rm max}$, the formula for the average travel time becomes
\begin{equation}
 \frac{w\cdot [(ax_1 + bx_1{}^2/2)-(ax_0 + bx_0{}^2/2)]}{ w\cdot (x_1-x_0)}
 = a + b\frac{x_1+x_0}{2}\, ,
\end{equation}
where we have used $(x_1{}^2 - x_0{}^2) = (x_1-x_0)(x_1+x_0)$.
Inserting the above parameters, we obtain the previously derived result
\begin{equation}
{\cal T}_i  = {\cal T}_i^0 +  \frac{\Delta N_i^{\rm max}}{2u_i\widehat{Q}_i} \, .
\end{equation}
When determining the average velocity $V_i^{\rm av}$, the function to average over is of the form
$f(x) =c/(a+bx)$, where $c=L_i$ and the other parameters are as defined before. We use the relationship
\begin{eqnarray}
\int\limits_{x_0}^{x_1} dx' \; \frac{wc}{a+bx'} &=& \frac{wc}{b} \Big( \ln |a+bx_1| - \ln |a+bx_0|\Big) \nonumber \\
&=& \frac{wc}{b} \ln \left| \frac{a+bx_1}{a+bx_0}\right| 
\end{eqnarray} 
Dividing this again by the normalization factor $w \cdot (x_1-x_0)$ and inserting the above parameters finally gives
\begin{equation}
	V_i^{\rm av} = \frac{L_iu_i\widehat{Q}_i}{\Delta N_i^{\rm max}} \ln \left| 1 + \frac{\Delta N_i^{\rm max}}{u_i\widehat{Q}_i{\cal T}_i^0}\right| 
	\approx \frac{L_i}{{\cal T}_i^0} \left(1 - \frac{\Delta N_i^{\rm max}}{2u_i\widehat{Q}_i{\cal T}_i^0}\right) \, ,
	\label{taylo1}
\end{equation}
where we have used $\ln(1+x) \le x - x^2/2$. 
This formula corrects the naive formula 
\begin{equation}
V_i^{\rm av} \approx \frac{L_i}{{\cal T}_i} = \frac{L_i}{{\cal T}_i^0 +  \frac{\Delta N_1^{\rm max}}{2u_i\widehat{Q}_i}}\approx \frac{L_i}{{\cal T}_i^0}\left( 1 - \frac{\Delta N_1^{\rm max}}{2u_i\widehat{Q}_i{\cal T}_i^0}\right)\, ,\label{taylo2}
\end{equation}
where we have used $1/(1+x) \approx 1 - x$. Therefore, the above Taylor approximations of both formulas agree, but higher-order approximations would differ. The formulas in the main part of the paper result for $N_i^{\rm av} = N_i^{\rm max}/2$, which  corresponds to the case $\delta_i = 0$ (i.e. $f_i - u_i$). 
\par
Generalizing the above approach to the case $\delta_i > 0$, we must split up the integrals into one over $wc/(a+bx')$ extending from $x_0=0$ to $x_1 = \Delta N_i^{\rm max}$ and another one over $wc/a$ from $x_1 = \Delta N_i^{\rm max}$ to $x_2 = (1-u_i)A_iT^{\rm cyc}= (1-u_i)u_i\widehat{Q}_i T^{\rm cyc}$, where the specifications of $a$, $b$, and $c$ are unchanged. Taking into account $V_i^0 = L_i/{\cal T}_i^0$, this gives
\begin{equation}
V_i^{\rm av} = \frac{wL_iu_i\widehat{Q}_i \ln | 1 + \Delta N_i^{\rm max}/(u_i\widehat{Q}_i{\cal T}_i^0)| + Z} {w(1-u_i) u_i\widehat{Q}_i T^{\rm cyc}} \, , 
\end{equation}
where
\begin{equation}
Z = wV_i^0 [ (1-u_i)u_i\widehat{Q}_i T^{\rm cyc} - \Delta N_i^{\rm max}] \, .
\end{equation}
Considering Eq. (\ref{dnmax}), we get 
\begin{equation}
V_i^{\rm av} = \frac{L_i}{(1-u_i)T_{\rm cyc}}\ln \left( 1 + (1-f_i) \frac{T_{\rm cyc}}{{\cal T}_i^0}\right)
+ V_i^0 \frac{f_i-u_i}{1-u_i} \, . \label{Taking}
\end{equation}
In second-order Taylor approximation, this results in 
\begin{equation}
V_i^{\rm av} \approx  V_i^0 \left[ \frac{1-f_i}{1-u_i} \left( 1 - \frac{(1-f_i)T_{\rm cyc}}{2{\cal T}_i^0} \right) + \frac{f_i-u_i}{1-u_i}\right] \, , 
\end{equation}
which can also be derived from Eq. (\ref{taylo2}), considering Eq. (\ref{dnmax}) and the percentage of delayed vehicles, which is given by Eq. (\ref{perca}). The same result follows from 
\begin{equation}
V_i^{\rm av} = \frac{L_i}{{\cal T}_i^0 + {\cal T}_i^{\rm av}} \approx V_i^0 \left( 1 - \frac{{\cal T}_i^{\rm av}}{{\cal T}_i^0} \right)
\end{equation}
together with Eq. (\ref{avtime1}). 

\begin{thebibliography}{99}
\bibitem{Gazis}
D. C. Gazis, {\it Traffic Theory} (Kluwer Academic, Boston, 2002).

\bibitem{Esser}
J.~Esser and M.~Schreckenberg, {Microscopic simulation of urban traffic
  based on cellular automata.}  {\em Int. J. Mod. Phys. B} {\bf 8},
  1025--1036 (1997).
  
\bibitem{Nagel1}
P. M. Simon and K. Nagel,
Simplified cellular automaton model for city traffic.
{\it Phys. Rev. E} {\bf 58}, 1286--1295 (1998).

\bibitem{Nagel} 
K. Nagel,  {\it Multi-Agent Transportation Simulations}, see 
http://www2.tu-berlin.de/fb10/ISS/FG4/archive/sim-archive/publications/book/

\bibitem{Hilliges} M. Hilliges and W. Weidlich,
A phenomenological model for dynamic traffic flow in networks.
{\it Transpn. Res. B} {\bf 29}, 407--431 (1995).

\bibitem{NHM}
D.~Helbing, J.~Siegmeier, and S.~L\"{a}mmer, {Self-organized network
  flows.}  {\em Networks and {H}eterogeneous {M}edia} {\bf 2}, 
  193--210 (2007).

\bibitem{Cremer} M. Cremer and J. Ludwig,
A fast simulation model for traffic flow on the basis of Boolean operations.
{\it Math. Comput. Simul.} {\bf 28}, 297ff (1986).

\bibitem{Daganzo}
C.~F. Daganzo, {The cell transmission model. {II}: {N}etwork traffic.}
  {\em Transpn. Res. {B}} {\bf 29} (1995), 79--93 (1995).

\bibitem{NagaCity}
T. Nagatani, 
Jamming transition in the traffic-flow model with two-level crossings.
{\it Phys. Rev. E}{\bf 48}, 3290-3294 (1993).


\bibitem{Chowdhury}
D.~Chowdhury and A.~Schadschneider, {Self-organization of traffic jams in
  cities: {E}ffects of stochastic dynamics and signal periods.}  {\em Phys.
  Rev. E} {\bf 59}, R1311--R1314 (1999).

\bibitem{BML} O. Biham, A. A. Middleton, and D. Levine,
Self-organization and a dynamical transition in traffic-flow models.
{\it Phys. Rev. A} {\bf 46}, R6124--R6127 (1992).

\bibitem{Zheng2007}
J.-F. Zheng, Z.-Y. Gao, and X.-M. Zhao, {Modeling cascading failures in
  congested traffic and transportation networks.}  {\em Phys. Stat. Mech.
  Appl.} {\bf 385}, 700--706 (2007).

\bibitem{Irwin}
N. A. Irwin, M. Dodd, and H. G. Von Cube,
Capacity restraint in multi-travel model assignment programs.
{\it Highway Research Board Bulletin} {\bf 347}, 258--289 (1961).

\bibitem{Smock}
R. J. Smock,
An iterative assignment approach to capacity restraint on arterial networks,
{\it Highway Research Board Bulletin} {\bf 347}, 60--66 (1962).

\bibitem{Mosher}
W. W. Mosher,
A capacity restraint algorithm for assignment flow to a transport network.
{\it Highway Research Record} {\bf 6}, 41--70 (1963).

\bibitem{BPR}
Bureau of Public Roads, {\it Traffic Assignment Manual}
(U.S. Dept. of Commerce, Urban Planning Division, Washington, D.C., 1964).

\bibitem{Soltmann}
T. J. Soltmann,
Effects of alternate loading sequences on results from Chicago trip distribution and assignment model.
{\it Highway Research Record} {\bf 114}, 122--140 (1965).

\bibitem{Davidson}
K. B. Davidson,
A flow travel-time relationship for use in transportation planning.
In: {\it Proceedings of the 3rd ARRB Conference}, Part 1
(Australian Road Research Board, Melbourne, 1966), pp. 183--194.

\bibitem{Smeed}
R. J. Smeed, 
Road capacity of city centers.
{\it Traffic Engineering and Control} {\bf 8}, 455--458 (1966).

\bibitem{Over}
K. R. Overgaard,
Urban transportation planning traffic estimation.
{\it Traffic Quarterly}, 197--218 (1967).

\bibitem{Thomson}
J. M. Thomson,
Speeds and flows in central London: 2. Speed-flow relations.
{\it Traffic Engineering and Control} {\bf 8}, 721--725 (1967).

\bibitem{Wardrop}
J. G. Wardrop, 
Journey speed and flow in central urban areas.
{\it Traffic Engineering and Control} {\bf 9}, 528--532 (1968).

\bibitem{Zahavi}
Y. Zahavi,
Traffic performance evaluation of road networks by the $\alpha$-relationship, Parts I and II.
{\it Traffic Engineering and Control} {\bf 14}, 228--231 and 292--293 (1972).

\bibitem{Akcelik}
R. Akcelik,
Travel time functions for transport planning purpose: Davidson's function, it's time-dependent
form and an alternative travel time function.
{\it Australian Road Research} {\bf 21}, 49--59 (1991).

\bibitem{Lum}
K. M. Lum, H. S. L. Fan, S. H. Lam, and P. Olszewski,
Speed-flow modeling of arterial roads in Singapore.
{\it J. Transpn. Eng.} {\bf 124}, 213--222 (1998).

\bibitem{Zhang}
H. M. Zhang, 
Link-journey-speed model for arterial traffic.
{\it Transpn. Res. Rec.} {\bf 1676}, 109--115 (1999).

\bibitem{Tu}
H. Tu, {\it Monitoring Travel Time Reliability on Freeways}
(Ph.D. thesis, Delft University of Technology, 2008).

\bibitem{Daganzo2007}
C.~F. Daganzo, {\it Urban gridlock: {M}acroscopic modeling and mitigation
  approaches},  {\em Transport. {R}es. {B}} {\bf 41}(1), 49-62 (2007).

\bibitem{Geroliminis}
N. Geroliminis and C. F. Daganzo,
Existence of urban-scale macroscopic fundamental diagrams: Some experimental findings.
{\it Transpn. Res. B} {\bf 42}, 759--770 (2008).

\bibitem{Simdata} N. Geroliminis and C. F. Daganzo,
Macroscopic modeling of traffic in cities.
TRB 86th Annual Meeting, Paper \#07-0413, Washington D.C. (2007).

\bibitem{DagGer}
C. F. Daganzo and N. Geroliminis,
An analytical approximation for the macroscopic fundamental diagram of urban traffic.
Accepted for publication (2008).

\bibitem{Godfrey}
J. W. Godfrey, 
The mechanism of a road network.
{\it Traffic Engineering and Control} {\bf 11}(7), 323--327 (1969).

\bibitem{twofluid} 
R. Herman and I. Prigogine,
A two-fluid approach to town traffic.
{\it Science} {\bf 204}, 148--151 (1979).

\bibitem{twofluid1}
R. Herman and S. Ardekani,
Charakterizing traffic conditions in urban areas.
{\it Transportation Science} {\bf 18}(2), 101-139 (1984).

\bibitem{Eichler} 
M. Eichler and C. F. Daganzo,
Bus lanes with intermittend priority: Strategy formulae and an evaluation.
{\it Transportation Research B} {\bf 40}(9), 731--744 (2006).

\bibitem{Review} D. Helbing, Traffic and related self-driven many-particle systems. {\it Reviews of Modern Physics} {\bf 73}, 1067-1141 (2001).

\bibitem{EPJBphase}
D. Helbing, M. Treiber, A. Kesting, and M. Sch\"onhof,
Theoretical vs. empirical classification and prediction of congested traffic states,
{\it European Physical Journal B}, submitted (2008).

\bibitem{Whitham} G. B. Whitham, {\it Linear and Nonlinear Waves} (Wiley, New

\bibitem{Little} 
R. Hall, {\it Queueing Methods for Service and Manufacturing} (Prentice Hall, Upper Saddle
River, NJ, 1991).
York, 1974).

\bibitem{JPhysA} D. Helbing, A section-based queueing-theoretical traffic model for congestion and travel time analysis in networks. {\it Journal of Physics A: Mathematical and General} {\bf 36}, L593-L598 (2003).

\bibitem{slower} D. Helbing, 
Operation regimes and slower-is-faster effect in the control of traffic intersections.
{\it European Journal of Physics B}, submitted (2008).

\bibitem{movinggaps} D. Helbing, T. Seidel, S. L\"ammer, and K. Peters, Self-organization principles in supply networks and production systems, in: {\it Econophysics and Sociophysics}, edited by B. K. Chakrabarti, A. Chakraborti, A. Chatterjee (Wiley, Weinheim, 2006), p. 552.

\end{thebibliography}
\end{document}